\documentclass[conference]{IEEEtran}
\usepackage{subcaption}
\usepackage{amsmath}
\usepackage{multirow}
\usepackage[hyphens]{url}
\usepackage{epsfig}
\usepackage[hidelinks]{hyperref}

\begin{document}

\date{}

\title{Deciphering Malware's use of TLS (without Decryption)}

\author{\IEEEauthorblockN{Blake Anderson}
\IEEEauthorblockA{Cisco\\
blaander@cisco.com}
\and
\IEEEauthorblockN{Subharthi Paul}
\IEEEauthorblockA{Cisco\\
subharpa@cisco.com}
\and
\IEEEauthorblockN{David McGrew}
\IEEEauthorblockA{Cisco\\
mcgrew@cisco.com}}

%\author{Blake Anderson \and Subharthi Paul \and David McGrew}
%
%\institute{Cisco Systems, Inc.\\
%\email{\{blaander,subharpa,mcgrew\}@cisco.com}
%}

\maketitle

\begin{abstract}

The use of TLS by malware poses new challenges to network threat detection because traditional pattern-matching techniques can no longer be applied to its messages. However, TLS also introduces a complex set of observable data features that allow many inferences to be made about both the client and the server. We show that these features can be used to detect and understand malware communication, while at the same time preserving the privacy of benign uses of encryption. These data features also allow for accurate malware family attribution of network communication, even when restricted to a single, encrypted flow.

To demonstrate this, we performed a detailed study of how TLS is used by malware and enterprise applications. We provide a general analysis on millions of TLS encrypted flows, and a targeted study on 18 malware families composed of thousands of unique malware samples and ten-of-thousands of malicious TLS flows. Importantly, we identify and accommodate the bias introduced by the use of a malware sandbox.   The performance of a malware classifier is correlated with a malware family's use of TLS, i.e., malware families that actively evolve their use of cryptography are more difficult to classify. 

We conclude that malware's usage of TLS is distinct from benign usage in an enterprise setting, and that these differences can be effectively used in rules and machine learning classifiers.

\end{abstract}

\section{Introduction}

Encryption is necessary to protect the privacy of end users. In a network setting, Transport Layer Security (TLS) is the dominant protocol to provide encryption for network traffic. While TLS obscures the plaintext, it also introduces a complex set of observable parameters that allow many inferences to be made about both the client and the server.

Legitimate traffic has seen a rapid adoption of the TLS standard over the past decade, with some studies stating that as much as 60\% of network traffic uses TLS \cite{encryptedTraffic}. Unfortunately, malware has also adopted TLS to secure its communication. In our dataset, $\sim$10\% of the malware samples use TLS. This trend makes threat detection more difficult because it renders the use of deep packet inspection (DPI) ineffective. It is important to determine whether encrypted network traffic is benign or malicious, and do so in a way that preserves the integrity of the encryption. And while 10\% of malware samples utilizing TLS seems low, we make the assumption that this number will increase as the level of encryption in network traffic increases. Along these lines, we have seen a slight, but statistically significant, increase in malicious, encrypted traffic over the past 12 months.

%The increase in malware's use of TLS is particularly troubling because most security products currently treat 

%To further motivate the need for a study exposing malware's use of TLS, we analyzed Snort \cite{snort}. Snort is a popular and freely available malware detection engine that relies on rules. As of this writing, there were 3,437 rules in the community rule set \cite{snortcommunity}. Only 48 rules were TLS specific, and of those, only 6 detected malware, all using strings in self-signed certificates. Of the remainder, 19 detect Heartbleed or other overflow attacks against TLS implementations, and 23 detect plaintext over ports typically assigned to TLS. This indicates that traditional signature-based techniques have not heavily invested in TLS-specific signatures. It is our goal in this paper to identify features and illustrate methodologies that allow for the creation of machine learning classifiers and, more generally, rules that can detect malicious, encrypted network communication. For instance, we identify features of both the TLS client and server gathered from unencrypted handshake messages that could be used to create snort rules.

To further motivate the need for a study exposing malware's use of TLS, we consider the limitations of a pattern-matching approach when faced with TLS, and analyzed a popular community Intrusion Protection System (IPS) rule set \cite{snortcommunity}. As of this writing, there were 3,437 rules in that set, 3,307 of which inspect packet contents. Only 48 rules were TLS specific, and of those, only 6 detected malware, using strings in self-signed certificates. Of the remainder, 19 detect Heartbleed or other overflow attacks against TLS implementations, and 23 detect plaintext over ports typically assigned to TLS. These numbers show that traditional signature-based techniques have not heavily invested in TLS-specific malware signatures to date. However, the rules that match certificate strings hint that it is possible to detect malware through the passive inspection of TLS. Our goal in this paper to confirm and substantiate this idea, by identifying data features and illustrating methodologies that allow for the creation of rules and machine learning classifiers that can detect malicious, encrypted network communication. For instance, we identify features of both the TLS client and server gathered from unencrypted handshake messages that could be used to create IPS rules.

In this paper, we provide a comprehensive study of malware's use of TLS by observing the unencrypted TLS handshake messages. We give a high-level overview of malware's use of TLS compared to what we have observed on an enterprise network. Enterprise traffic typically uses up-to-date cryptographic parameters that are indicative of up-to-date TLS libraries. On the other hand, malware typically uses older and weaker cryptographic parameters. Malware's usage of TLS is distinct compared to enterprise traffic, and, for most families, this fact can be leveraged to accurately classify malicious traffic patterns. We examine these difference from both a TLS client and a TLS server perspective.

%As an example of a commonly used and freely available malware detection engine, we analyzed Snort \cite{snort} and its community rule set \cite{snortcommunity}. As of this writing, there were 3,437 rules in the community rule set. 48 of these rules were SSL/TLS specific: 17 to detect Heartbleed, 6 to detect self-signed certificates with various parameters, and 2 to detect overflow attempts against particular SSL implementations. The remaining 23 rules detected plaintext over ports typically assigned to TLS. This indicates that traditional signature-based techniques have not heavily invested in TLS-specific signatures. It is our goal in this paper to identify features and illustrate methodologies that allow for the creation of machine learning classifiers and rules that can detect malicious, encrypted network communication. For instance, we identify features of both the TLS client and server gathered from unencrypted handshake messages that could be used to create snort rules.

In addition to our in-depth technical analysis, it is interesting to note the general tone that malware authors have towards encryption. There is an FAQ section in the open-sourced Zeus/Zbot malware \cite{zeusSource} where the following question and answer occur (content left as is):

\hspace{3mm} \textbf{Question:} \textit{Why traffic is encrypted with symmetric encryption method (RC4), but not asymmetric (RSA)?}

\hspace{3mm} \textbf{Answer:} \textit{Because, in the use of sophisticated algorithms it makes no sense, encryption only needs to hide traffic.}

\noindent In the current privacy climate, this attitude most certainly does not hold for enterprise network traffic \cite{weakdh15,opderbeck2016apple}. Again, this divergence is another tool we can take advantage of to more accurately classify malicious flows.

When applying machine learning classifiers on a per-family basis, it is clear that some families/subfamilies are more difficult to classify. Our goal is not to show optimized machine learning classifiers, but rather to identify what characteristics of the specific family make it difficult to classify. For instance, we can correlate poor classifier performance on encrypted traffic patterns with one family's use of strong \cite{strongCrypto} and varied cryptography. We also examine additional features extracted from unencrypted TLS handshake messages that significantly increase the performance of the classifiers. In general, we have found this approach to be fruitful: identify weaknesses in the features used to represent a flow on a per-family basis, and then augment that representation with more informative features.

Finally, we show how we can perform family attribution given only network based data. This problem is positioned as a multi-class classification problem where each malware family has its own label. We identify families who use identical TLS parameters, but can still be accurately classified because their traffic patterns with respect to other flow-based features are distinct. We also identify subfamilies of malware that cannot be distinguished from one another with only their network data. We are able to achieve an accuracy of 90.3\% for the family attribution problem when restricted to a single, encrypted flow, and an accuracy of 93.2\% when we make use of all encrypted flows within a 5-minute window.

We use a commercial sandbox environment to collect the first five minutes of a malware sample's network activity. We collected tens-of-thousands of unique malware samples and hundreds-of-thousands of malicious, encrypted flows from these samples. We collected millions of TLS encrypted flows from an enterprise network to compare against the malware data. We used an open source project to collect the data and transform it to a JSON format that contained the typical network 5-tuple, the sequence of packet lengths and inter-arrival times, the byte distribution, and the unencrypted TLS handshake information. All of the analysis done in this paper uses only network data, and does not assume an endpoint presence.

The remainder of the paper is organized as follows: Section \ref{sec:assumptions} outlines some basic assumptions we make with respect to the data and our methodology, and Section \ref{sec:data} reviews how we obtained our data, specifies the datasets we use for each experiment, and what features we use to classify the network flows. Section \ref{sec:malware_families_and_tls} gives an overview of how malware's usage of TLS differs from that of an enterprise network from both the perspective of a TLS client and a TLS server. Section \ref{sec:malware_classification} shows which families are difficult to classify from a network flow point-of-view, and explains why this is the case, and Section \ref{sec:family_attribution} gives results showing how we can attribute a flow to a particular family. Section \ref{sec:related_work} reviews previous and related work, Section \ref{sec:limitations} outlines some limitations of our approach, and finally, we conclude in Section \ref{sec:conclusions}.

\begin{table}[t!]\small%\normalsize
\center
  \begin{tabular}{l|r}
  	\hline
    \hspace{.5mm}Port\hspace{.5mm} & \hspace{.5mm}Percentage of TLS Flows\hspace{.5mm} \\
    \hline
    \hline
    443 & 98.4\% \\
    \hline
    9001 & 1.2\% \\
    \hline
    80 & 0.1\% \\
    \hline
    9101 & 0.1\% \\
    \hline
    9002 & 0.1\% \\
    \hline
  \end{tabular}
  \vspace{2mm}
  \caption{Based on malware data collected between August 2015 and May 2016, we investigated which ports malware used the most for TLS Encrypted communication.}
  \label{table:tls_ports}
\end{table}

\section{Preliminaries and Assumptions}
\label{sec:assumptions}

Our primary concern in this paper is to categorize and classify malicious, TLS encrypted flows. While we do use the \texttt{serverHello} and \texttt{certificate} messages to highlight some interesting features about the servers that the malware samples are connecting to, our main focus is client oriented. The classification algorithms we develop are heavily dependent on client-based features, which allows our algorithms to correctly classify a malicious agent connecting to \texttt{google.com} versus a typical enterprise agent connecting to \texttt{google.com}, i.e., we can leverage the client's cryptographic parameters to differentiate these two events. For this reason, we do not filter the malware's TLS traffic to only include command and control flows, but also allow other types of TLS-encrypted traffic such as click-fraud.

In this paper, we focus  on TLS encrypted flows over port 443 to make the comparisons between enterprise TLS and malicious TLS be as unbiased as possible. To further motivate this choice, Table \ref{table:tls_ports} lists the 5 most used ports for TLS by the malware samples collected between August 2015 and May 2016. To determine if a flow was TLS, we used deep packet inspection and a custom signature based on the TLS versions and message types of the \texttt{clientHello} and \texttt{serverHello} messages. In total, we found 229,364 TLS flows across 203 unique ports, and port 443 was by far the most common port for malicious TLS. Although the diversity of port usage in malware was great, these diverse ports were relatively uncommon.

Given that our non-malware data was collected on an enterprise network, it naturally follows that the categorization and classification results presented in this paper are most applicable to the enterprise setting. We do not claim that these results hold for the general class of networks, e.g., service provider data. That being said, we do believe that securing enterprise networks is an important use case and that the conclusions presented in this paper offer enterprise network operators significantly novel and valuable results.

The enterprise network data used in this paper was initially filtered using a well known IP blacklist \cite{talos}. This removed $\sim$0.05\% of the initial traffic. After this filtering stage, we take the data ``as-is". We are aware that there is most likely more malicious traffic in this dataset, but this fact is just taken as a base assumption for reasons of practicality.

%we are concerned about classifying malicious flows, not malicious servers, while we do highlight some interesting features about server certs and parameters, our main focus is more client oriented. classification uses client-based features. client behavior - malware google.com versus firefox google.com. 

%\begin{table*}[t!]\normalsize
%\center
%  \begin{tabular}{|l|r|r|r|r|r|r|r|r|r|r|r|r|}
%  	\hline
%  	\multirow{2}{*}{Problem} & \multicolumn{10}{|c|}{Malware} & \multicolumn{2}{|c|}{Benign} \\
%  	%\hline
%     & Aug & Sept & Oct & Nov & Dec & Jan & Feb & Mar & Apr & May & May & June \\
%    \hline
%    \hline
%     &  &  &  &  &  &  &  &  &  &  &  & \\
%    \hline
%  \end{tabular}
%  \vspace{2mm}
%  \caption{}
%  \label{table:dataset_overview}
%\end{table*}

\section{Data}
\label{sec:data}

The data for this paper was collected from a commercial sandbox environment where users can submit suspicious executables. Each submitted sample is allowed to run for 5 minutes. The full packet capture is collected and stored for each sample. Due to constraints of the sandbox environment, all network traffic observed in the sandbox is considered to be that of the originally submitted sample. For instance, if sample $A$ downloads and installs $B$ and $C$, then the traffic generated from $B$ and $C$ would be considered $A$'s.

This method of data collection is straightforward, and while it ignores some details about what is occurring on the endpoint, it is consistent with our goal of understanding each sample based solely on its network communications. Some biases were introduced with this approach. First, to reduce the number of false positives, we only considered samples that were known bad. In this setting, known bad means hitting on four or more antivirus convictions from unique vendors in VirusTotal \cite{virustotal}. Second, due to hardware constraints, the samples are only allowed to run for 5 minutes in a Windows XP-based virtual machine. Any encrypted network traffic that happens after this initial 5 minute window will not be captured. Similarly, any samples that are not compatible with Windows XP will not run in this environment.

The enterprise data was collected from an enterprise network with $\sim$500 active users and $\sim$4,000 unique IP addresses. The majority of the machines on this network run Windows 7, with the second most popular operating system being OS X El Capitan.

%Benign Data *************** - 4 day period in June, 4 day period in May

\subsection{Dataset and Sample Selection}

\begin{table}[t!]\small%\normalsize
\center
  \begin{tabular}{l|r|r}
  	\hline
    \hspace{.5mm}Malware Family\hspace{.5mm} & \hspace{.5mm}Unique Samples\hspace{.5mm} & \hspace{.5mm}Encrypted Flows\hspace{.5mm} \\
    \hline
    \hline
    Bergat & 192 & 332 \\
    \hline
    Deshacop & 69 & 129 \\
    \hline
    Dridex & 38 & 103 \\
    \hline
    Dynamer & 118 & 372 \\
    \hline
    Kazy & 228 & 1,152 \\
    \hline
    Parite & 111 & 275 \\
    \hline
    Razy & 117 & 564 \\
    \hline
    Sality & 612 & 1,200 \\
    \hline
    Skeeyah & 81 & 218 \\
    \hline
    Symmi & 494 & 2,618 \\
    \hline
    Tescrypt & 137 & 205 \\
    \hline
    Toga & 156 & 404 \\
    \hline
    Upatre & 377 & 891 \\
    \hline
    Virlock & 1,208 & 12,847 \\
    \hline
    Virtob & 115 & 511 \\
    \hline
    Yakes & 100 & 337 \\
    \hline
    Zbot & 1,291 & 2,902 \\
    \hline
    Zusy & 179 & 733 \\
    \hline
    \hline
    Total & 5,623 & 25,793 \\
    \hline
  \end{tabular}
  \vspace{2mm}
  \caption{Summary of the malicious families used in our analysis. We collected 18 malicious families, 5,623 malicious samples, and 25,793 encrypted flows that successfully negotiated the TLS handshake and sent application data.}
  \label{table:malware_summary}
\end{table}

The malware traffic used in this paper was collected from August 2015 to May 2016, and the enterprise traffic was collected during a 4 day period in May 2016 and a 4 day period in June 2016. In this work, we performed several experiments on different subsets of this data. 

We first analyze the differences between the TLS parameters typically seen on an enterprise network versus the TLS parameters used by the general malware population. To proceed, we first removed all of the TLS flows that offered an ordered ciphersuite list that matched a list found in the default Windows XP \texttt{SChannel} implementation \cite{schannel}. This was done to help ensure that the TLS clients we observed were representative of the malware's behavior and not that of the TLS library provided by the underlying operating system. This removed $\sim$40\% of the malicious TLS flows and $\sim$0.4\% of the enterprise TLS flows. After this filtering stage, we used all of the TLS flows in our dataset. From August 2015 until May 2016, we collected 133,744 TLS flows initiated by malicious programs. During the 4 day periods in May and June 2016, we collected 1,500,005 TLS flows from an enterprise network. All of these TLS flows successfully negotiated the full TLS handshake and sent application data.

To analyze the differences between the TLS parameters used by different malware families, we used the malware samples from October 2015 to May 2016 that had an identifiable family name. Table \ref{table:malware_summary} gives a summary of the number of samples and flows for each malware family. The family name was generated by a majority vote from the signatures provided by VirusTotal \cite{virustotal}. Malware samples without a clear family name were discarded, i.e., any sample without at least four different antivirus programs using the same name (ignoring common names such as \texttt{Trojan}). Family names with less than 100 flows were not used. This process pruned our original set of 20,548 samples that used TLS to 5,623 unique samples across 18 families. It is difficult to determine the family, if any, associated with a malware sample, even with the information provided through dynamic analysis in a sandbox setting. These samples generated 25,793 TLS encrypted flows that successfully negotiated the TLS handshake and sent application data.

In this paper, we also make use of machine learning classifiers in three experiments. The first is to demonstrate the value of the additional TLS features through 10-fold cross-validation. For this experiment, we use all of the malicious TLS flows collected from August 2015 until May 2016, and a random subset of the May and June 2016 enterprise network's TLS flows. In total, there were 225,740 malicious and 225,000 enterprise flows for this experiment. To account for the bias that the Windows XP-based sandbox could introduce, we also present results on a dataset composed of only flows that offered an ordered ciphersuite list that did not match a list found in the default Windows XP \texttt{SChannel} implementation: 133,744 malicious and 135,000 enterprise TLS flows.

In the next set of experiments, we analyzed how well a trained classifier is able to detect the TLS flows generated by the different malware families. To train the classifier, we used the same 225,000 enterprise flows as above for the negative class, and 76,760 malicious TLS flows collected during August and September 2015 for the positive class. The testing data consisted of the TLS flows from October 2015 to May 2016 that could be assigned a ground truth family as described above. Again, Table \ref{table:malware_summary} gives a summary of the number of samples and flows for each malware family. While we do not remove flows that offered an ordered ciphersuite list that matched a list found in the default Windows XP \texttt{SChannel} implementation in this experiment, we do make explicit the families that have this bias.

Finally, to assess the malware family attribution potential of TLS handshake metadata, we used 10-fold cross-validation and multi-class classification on the data listed in Table \ref{table:malware_summary}. Again, we do not remove samples that offered an ordered ciphersuite list that matched a list found in the default Windows XP \texttt{SChannel} implementation in this experiment because all of the samples would have the same bias.

\subsection{Feature Extraction}

To extract the data features of interest, we wrote software tools to extract the data features of interest from live traffic or packet capture files. The open source project will export all of the data in a convenient JSON format. The machine learning classifiers are built using traditional flow features, traditional ``side-channel" features, and features collected from the unencrypted TLS handshake messages.

%\noindent
%\textbf{Flow Metadata.}

\subsubsection{Flow Metadata}

The first set of features investigated are modeled around traditional flow data that is typically collected in devices configured to export IPFIX/NetFlow. These features include the number of inbound bytes, outbound bytes, inbound packets, outbound packets; the source and destination ports; and the total duration of the flow in seconds. These features were normalized to have zero mean and unit variance.

%\begin{figure*}[t!]
%    \centering
%    \begin{subfigure}[b]{0.49\textwidth}
%           \centering
%           \includegraphics[scale=1.1]{figures/tls_benign.png}
%            \caption{TLS (Video)}
%            \label{fig:a}
%    \end{subfigure}
%    \begin{subfigure}[b]{0.49\textwidth}
%            \centering
%            \includegraphics[scale=1.1]{figures/tls_malware.png}
%            \caption{TLS (Malware C2)}
%            \label{fig:b}
%    \end{subfigure}
%    \caption{Two distinct, encrypted byte distributions. The byte distribution provides an interesting, additional data source for encrypted network traffic.}
%    \label{figure:byte_distribution}
%\end{figure*}

\subsubsection{Sequence of Packet Lengths and Times.}

The sequence of packet lengths and packet inter-arrival times (SPLT) has been well studied \cite{nguyen2008survey,zander2005automated}. In our open source implementation, the SPLT elements are collected for the first 50 packets of a flow. Zero-length payloads (such as ACKs) and retransmissions are ignored.

A Markov chain representation is used to model the SPLT data. For both the lengths and times, the values are discretized into equally sized bins, e.g., for the length data, 150 byte bins are used where any packet size in the range [0,150) will go into the first bin, any packet size in the range [150,300) will go into the second bin, etc. A matrix $A$ is then constructed where each entry, $A[i,j]$, counts the number of transitions between the $i$'th and $j$'th bin. Finally, the rows of $A$ are normalized to ensure a proper Markov chain. The entries of $A$ are then used as features to the machine learning algorithms.

\subsubsection{Byte Distribution.}

The byte distribution is a length-256 array that keeps a count for each byte value encountered in the payloads of the packets for each packet in the flow. The byte value probabilities can be easily computed by dividing the byte distribution counts by the total number of bytes found in the packets' payloads. The 256 byte distribution probabilities are used as features by the machine learning algorithms. The full byte distribution provides a lot of information about the encoding of the data. Additionally, the byte distribution can give information about the header-to-payload ratios, the composition of the application headers, and if any poorly implemented padding is added.

%\begin{figure*}[t!]
%	\centering
%   \includegraphics[scale=0.65]{figures/tls_client.pdf}
%   \caption{Malware's use of TLS versus that of enterprise network traffic relative to the TLS client features. Some values and the full ciphersuite names were omitted for clarity of presentation. Ciphersuites and extensions are represented as hex codes, which are given in full in Appendix \ref{app:hex_codes}.}
%   \label{figure:malware_versus_benign_client}
%
%	\vspace*{\floatsep}
%
%   \includegraphics[scale=0.65]{figures/tls_server.pdf}
%   \caption{Malware's use of TLS versus that of enterprise network traffic relative to the TLS server features. Some values and the full ciphersuite names were omitted for clarity of presentation. Ciphersuites and extensions are represented as hex codes, which are given in full in Appendix \ref{app:hex_codes}.}
%   \label{figure:malware_versus_benign_server}
%\end{figure*}

\subsubsection{Unencrypted TLS Header Information.}

TLS (Transport Layer Security) is a cryptographic protocol that provides privacy for applications. TLS is usually implemented on top of common protocols such as HTTP for web browsing or SMTP for email. HTTPS is the usage of TLS over HTTP, which is the most popular way of securing communication between a web server and client, and is supported by most major web servers. HTTPS typically uses port 443.

The TLS version, the ordered list of offered ciphersuites, and the list of supported TLS extensions are collected from the \texttt{client hello} message. The selected ciphersuite and selected TLS extensions are collected from the \texttt{server hello} message. The server's certificate is collected from the \texttt{certificate} message. The client's public key length is collected from the \texttt{client key exchange} message, and is the length of the RSA ciphertext or DH/ECDH public key, depending on the ciphersuite. Similar to the sequence of packet lengths and times, the sequence of record lengths, times, and types is collected from TLS sessions.

%\begin{figure*}[t!]
%	\centering
%   \includegraphics[scale=0.41]{figures/offered_ciphersuites.pdf}
%   \caption{Comparison of the ciphersuites benign traffic offers versus the ciphersuites that malware offers.}
%   \label{figure:offered_ciphersuites}
%\end{figure*}

In our classification algorithms, the list of offered ciphersuites, the list of advertised extensions, and the client's public key length were used. 176 offered ciphersuite hex codes were observed in our full dataset, and a binary vector of length 176 was created where a one is assigned to each ciphersuite in the list of offered ciphersuites. Similarly, we observed 21 unique extensions, and a binary vector of length 21 was created where a one is assigned to each extension in the list of advertised extensions. Finally, the client's public key length was represented as a single integer value. In total, 198 TLS client-based features were used in the classification algorithms. In some experiments, we use an additional TLS server-based binary feature: whether the certificate was self-signed or not.

\begin{figure*}[t!]
	\centering
   \includegraphics[scale=0.74]{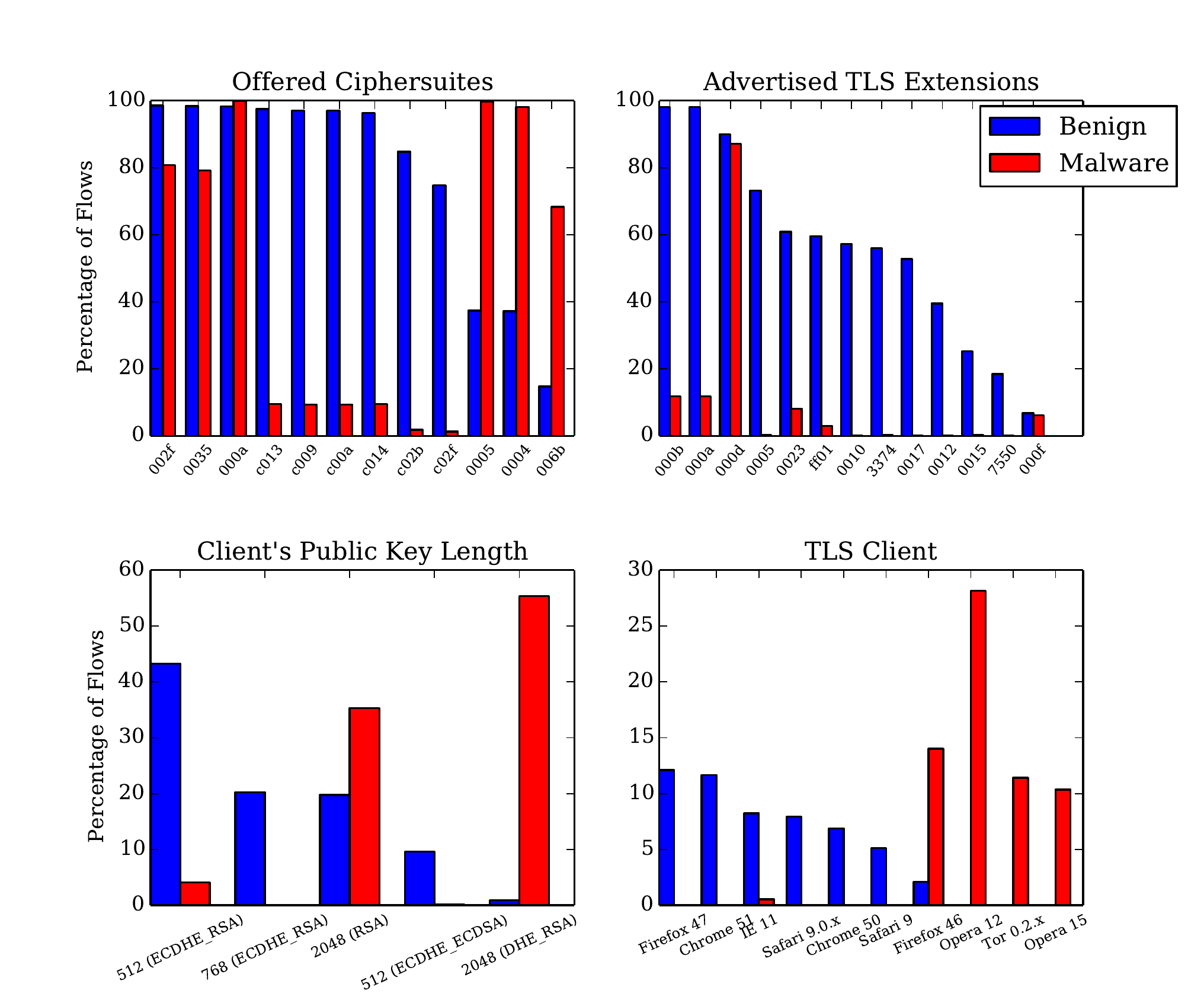}
   \caption{Malware's use of TLS versus that of enterprise network traffic relative to the TLS client features. Some values and the full ciphersuite names were omitted for clarity of presentation. Ciphersuites and extensions are represented as hex codes, which are given in full in Appendix \ref{app:hex_codes}.}
   \label{figure:malware_versus_benign_client}
\end{figure*}

\section{Malware Families and TLS}
\label{sec:malware_families_and_tls}

Although malware uses TLS to secure its communication, our data suggests that for the majority of the families we analyzed, malware's use of TLS is quite distinct from that of the enterprise network's traffic. In this section, we highlight these differences from the perspective of the TLS client and also from the perspective of the TLS server.

For the comparisons between general malware and enterprise traffic, we first removed all of the TLS flows that offered an ordered ciphersuite list that matched a list found in the default Windows XP \texttt{SChannel} implementation \cite{chooseSchannel,ssllabs} from our full dataset. We found that $\sim$40\% of TLS flows from malware samples offered this list. To help ensure that our analysis was capturing trends in the malware's use of TLS, and not that of the underlying operating system, we removed all of these flows. After this filtering stage, we used all of the TLS flows in our dataset. From August 2015 to May 2016, we collected 133,744 TLS flows initiated by malicious programs that successfully negotiated the full TLS handshake and sent application data. In May and June 2016, we collected 1,500,005 TLS flows from an enterprise network using the same criteria.

The malware data collection process can introduce biases in terms of malware family representation, and the conclusions that can be made from the TLS parameters collected. To account for this, we also analyze the TLS clients that malware uses and the TLS servers that malware connects to on a per-family basis. In this analysis, we highlight the families that use the default Windows TLS library, and the families which include their own TLS client. The data for this experiment is listed in Table \ref{table:malware_summary}.

\subsection{TLS Clients}

\subsubsection{Malware versus Enterprise}

Figure \ref{figure:malware_versus_benign_client} illustrates the differences between the malware's and the enterprise's usage of TLS with respect to the TLS clients after filtering typical Windows XP ciphersuite lists. Nearly 100\% of the enterprise TLS sessions offered the \texttt{0x002f} (\texttt{TLS\_RSA\_WITH\_AES\_128\_CBC\_SHA}) ciphersuite and the \texttt{0x0035} (\texttt{TLS\_RSA\_WITH\_AES\_256\_CBC\_SHA}) ciphersuite. On the other hand, nearly 100\% of the malicious TLS sessions observed offered:
\begin{itemize}
\item \texttt{0x000a} \newline(\texttt{TLS\_RSA\_WITH\_3DES\_EDE\_CBC\_SHA})
\item \texttt{0x0005} (\texttt{TLS\_RSA\_WITH\_RC4\_128\_SHA})
\item \texttt{0x0004} (\texttt{TLS\_RSA\_WITH\_RC4\_128\_MD5})
\end{itemize}
These three ciphersuites are considered weak, and although the enterprise traffic we observed does offer these ciphersuites, it does not offer them with the same frequency that the malicious traffic does.

\begin{table*}[t!]\small%\normalsize
\center
  \begin{tabular}{l|r|r|r|r|r}
  	\hline
    \multicolumn{1}{c|}{Malware} & \multicolumn{1}{|c|}{Number} & \multicolumn{1}{|c|}{Most Seen}  & \multicolumn{1}{|c|}{Number of Distinct} & \multicolumn{1}{|c|}{Most Frequently} & \multicolumn{1}{|c}{Client's} \\
    \multicolumn{1}{c|}{Family} & \multicolumn{1}{|c|}{of Flows} & \multicolumn{1}{|c|}{TLS Client} & \multicolumn{1}{|c|}{Ciphersuite Offer Vectors} & \multicolumn{1}{|c|}{Advertised Extension} & \multicolumn{1}{|c}{Public Key} \\
    \hline
    \hline
    Bergat & 332 & \texttt{IE 8}* & 1 & \texttt{None} & 2048-bit (\texttt{RSA}) \\
    \hline
    Deshacop & 129 & \texttt{Tor Browser 4} & 3 & \texttt{SessionTicket TLS} & 2048-bit (\texttt{RSA}) \\
    \hline
    Dridex & 103 & \texttt{IE 11} & 5 & \texttt{ec\_point\_formats} & 2048-bit (\texttt{RSA}) \\
     &  &  & & \texttt{supported\_groups} &  \\
     &  &  & & \texttt{renegotiation\_info} &  \\
    \hline
    Dynamer & 372 & \texttt{Tor 0.2.2} & 10 & \texttt{SessionTicket TLS} & 512-bit (\texttt{ECDHE\_RSA}) \\
    \hline
    Kazy & 1152 & \texttt{IE 8}* & 5 & \texttt{None} & 2048-bit (\texttt{RSA}) \\
    \hline
    Parite & 275 & \texttt{IE 8}* & 11 & \texttt{None} & 2048-bit (\texttt{RSA}) \\
    \hline
    Razy & 564 & \texttt{Tor Browser 4} & 8 & \texttt{None} & 2048-bit (\texttt{RSA}) \\
    \hline
    Sality & 1,200 & \texttt{IE 8}* & 133 & \texttt{None} &  2048-bit (\texttt{RSA}) \\
    \hline
    Skeeyah & 218 & \texttt{Tor 0.2.7} & 11 & \texttt{SessionTicket TLS} &  512-bit (\texttt{ECDHE\_RSA})\\
    \hline
    Symmi & 2,618 & \texttt{Opera 15} & 19 & \texttt{ec\_point\_formats} & 512-bit (\texttt{ECDHE\_RSA}) \\
     &  &  &  & \texttt{supported\_groups} &  \\
    \hline
    Tescrypt & 205 & \texttt{IE 8}* & 6 & \texttt{None} & 2048-bit (\texttt{RSA}) \\
    \hline
    Toga & 404 & \texttt{Tor 0.2.2} & 2 & \texttt{SessionTicket TLS} & 2048-bit (\texttt{RSA}) \\
     &  &  &  & \texttt{ec\_point\_formats} &  \\
     &  &  &  & \texttt{supported\_groups} &  \\
    \hline
    Upatre & 891 & \texttt{IE 8}* & 3 & \texttt{None} & 2048-bit (\texttt{RSA}) \\
    \hline
    Virlock & 12,847 & \texttt{Opera 12} & 1 & \texttt{signature\_algorithms} & 2048-bit (\texttt{DHE\_RSA}) \\
    \hline
    Virtob & 511 & \texttt{IE 8}* & 4 & \texttt{None} & 2048-bit (\texttt{RSA}) \\
    \hline
    Yakes & 337 & \texttt{IE 8}* & 3 & \texttt{None} & 2048-bit (\texttt{RSA}) \\
    \hline
    Zbot & 2,902 & \texttt{IE 8}* & 12 & \texttt{None} & 2048-bit (\texttt{RSA}) \\
    \hline
    Zusy & 733 & \texttt{IE 8}* & 7 & \texttt{None} & 2048-bit (\texttt{RSA}) \\
    \hline
  \end{tabular}
  \vspace{2mm}
  \caption{The most popular TLS client configurations for the 18 malicious families. The TLS client was estimated using TLS fingerprinting techniques \cite{ssllabs}. For TLS extensions, in the case of a tie, all equally probable extensions are listed. (*) indicates the fingerprint of the TLS client provided by the underlying sandbox operating system.}
  \label{table:malware_tls_client}
\end{table*}

\begin{figure*}
	\centering
   \includegraphics[scale=0.74]{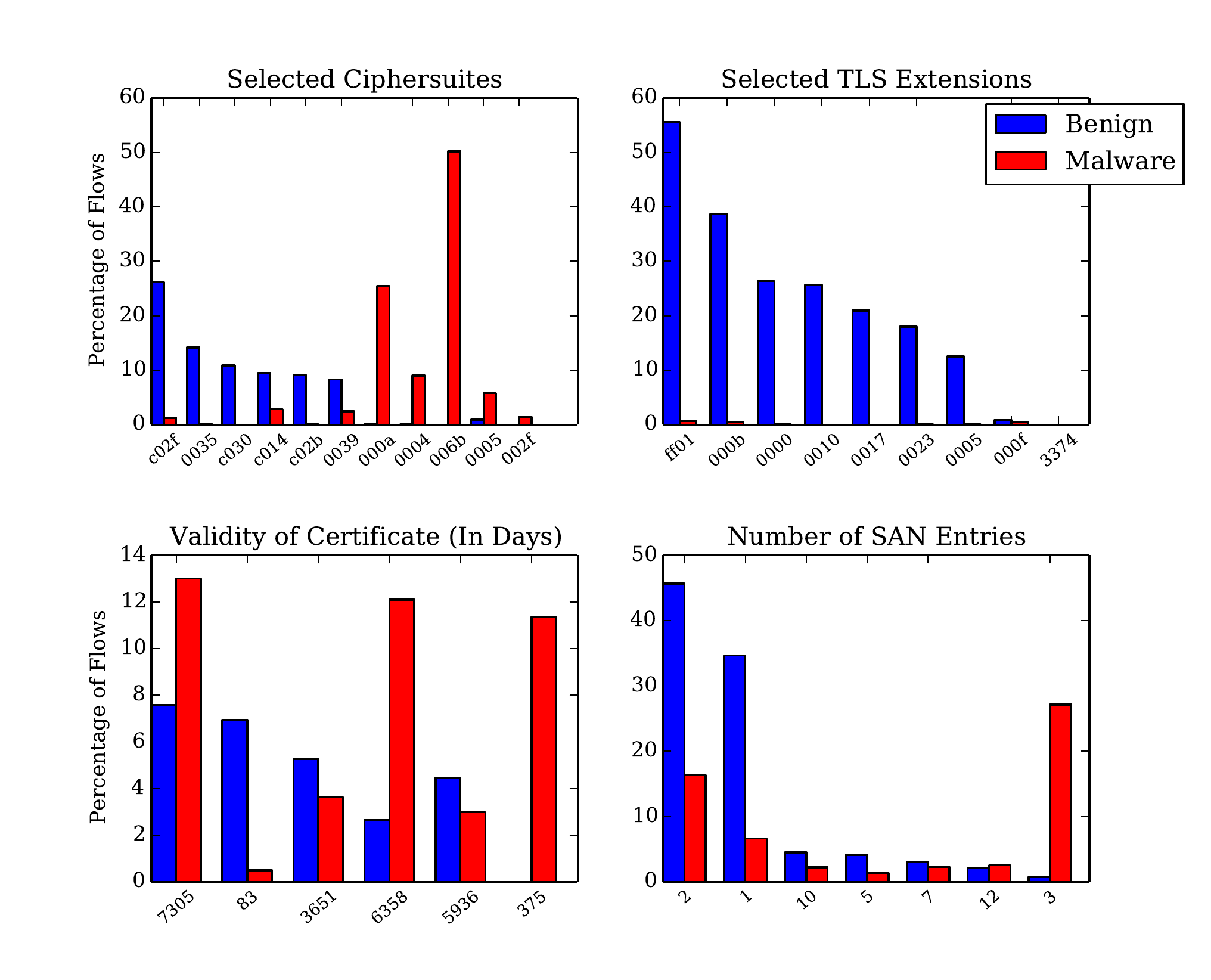}
   \caption{Malware's use of TLS versus that of enterprise network traffic relative to the TLS server features. Some values and the full ciphersuite names were omitted for clarity of presentation. Ciphersuites and extensions are represented as hex codes, which are given in full in Appendix \ref{app:hex_codes}.}
   \label{figure:malware_versus_benign_server}
\end{figure*}

\begin{table*}[t!]\small%\normalsize
\center
  \begin{tabular}{l|r|r|r|r|r}
  	\hline
    \multicolumn{1}{c|}{Malware} & \multicolumn{1}{|c|}{Number} & \multicolumn{1}{|c|}{Unique} & \multicolumn{1}{|c|}{Number of} & \multicolumn{1}{|c|}{Selected} & \multicolumn{1}{|c}{Certificate} \\
    \multicolumn{1}{c|}{Family} & \multicolumn{1}{|c|}{of Flows} & \multicolumn{1}{|c|}{Server IPs} & \multicolumn{1}{|c|}{SS Certs} & \multicolumn{1}{|c|}{Ciphersuite} & \multicolumn{1}{|c}{Subject} \\
    \hline
    \hline
    Bergat & 332 & 12 & 0 & \texttt{TLS\_RSA\_WITH\_3DES\_EDE\_CBC\_SHA} &  \texttt{www.dropbox.com} \\
    \hline
    Deshacop & 129 & 38 & 0 & \texttt{TLS\_RSA\_WITH\_3DES\_EDE\_CBC\_SHA} &  \texttt{*.onion.to} \\
    \hline
    Dridex & 103 & 10 & 89 & \texttt{TLS\_RSA\_WITH\_AES\_128\_CBC\_SHA} &  \texttt{amthonoup.cy} \\
    \hline
    Dynamer & 372 & 155 & 3 & \texttt{TLS\_ECDHE\_RSA\_WITH\_AES\_128\_GCM\_SHA256} &  \texttt{www.dropbox.com} \\
    \hline
    Kazy & 1152 & 225 & 52 & \texttt{TLS\_RSA\_WITH\_3DES\_EDE\_CBC\_SHA} &  \texttt{*.onestore.ms} \\
    \hline
    Parite & 275 & 128  & 0 & \texttt{TLS\_RSA\_WITH\_3DES\_EDE\_CBC\_SHA} &  \texttt{*.google.com} \\
    \hline
    Razy & 564 & 118 & 16 & \texttt{TLS\_RSA\_WITH\_RC4\_128\_SHA} &  \texttt{baidu.com} \\
    \hline
    Sality & 1,200 & 323 & 4 & \texttt{TLS\_RSA\_WITH\_3DES\_EDE\_CBC\_SHA} &  \texttt{vastusdomains.com} \\
    \hline
    Skeeyah & 218 & 90 & 0 & \texttt{TLS\_ECDHE\_RSA\_WITH\_AES\_128\_GCM\_SHA256} &  \texttt{www.dropbox.com} \\
    \hline
    Symmi & 2,618 & 700 & 22 & \texttt{TLS\_ECDHE\_RSA\_WITH\_AES\_256\_CBC\_SHA} &  \texttt{*.criteo.com} \\
    \hline
    Tescrypt & 205 & 26 & 0 & \texttt{TLS\_RSA\_WITH\_3DES\_EDE\_CBC\_SHA} &  \texttt{*.onion.to} \\
    \hline
    Toga & 404 & 138 & 8 & \texttt{TLS\_RSA\_WITH\_3DES\_EDE\_CBC\_SHA} &  \texttt{www.dropbox.com} \\
    \hline
    Upatre & 891 & 37 & 155 & \texttt{TLS\_RSA\_WITH\_RC4\_128\_MD5} &  \texttt{*.b7websites.net} \\
    \hline
    Virlock & 12,847 & 1 & 0 & \texttt{TLS\_DHE\_RSA\_WITH\_AES\_256\_CBC\_SHA256} &  \texttt{block.io} \\
    \hline
    Virtob & 511 & 120 & 0 & \texttt{TLS\_RSA\_WITH\_3DES\_EDE\_CBC\_SHA} &  \texttt{*.g.doubleclick.net} \\
    \hline
    Yakes & 337 & 51 & 0 & \texttt{TLS\_RSA\_WITH\_RC4\_128\_SHA} &  \texttt{baidu.com} \\
    \hline
    Zbot & 2,902 & 269 & 507 & \texttt{TLS\_RSA\_WITH\_RC4\_128\_MD5} & \texttt{tridayacipta.com} \\
    \hline
    Zusy & 733 & 145 & 14 & \texttt{TLS\_RSA\_WITH\_3DES\_EDE\_CBC\_SHA} & \texttt{*.criteo.com} \\
    \hline
  \end{tabular}
  \vspace{2mm}
  \caption{TLS server configurations for the servers most visited by the 18 malicious families. The certificate subject typically has a long tail, but only the most frequent is reported. The reported number of self-signed certificates is not necessarily related to the most popular certificate subject.}
  \label{table:malware_tls_server}
\end{table*}

The differences in malware and enterprise's TLS \texttt{client hello} messages become more evident when the advertised TLS extensions are considered. We observed a much greater diversity in the TLS extensions that enterprise clients advertised. Almost half of the enterprise clients would advertise up to 9 extensions, but the malicious clients would only consistently advertise one: \texttt{0x000d} (\texttt{signature\_algorithms}), an RFC MUST in most circumstances \cite{rfc5246}. The following four extensions were observed in $\sim$50\% of the enterprise traffic and rarely observed in the malicious traffic: 
\begin{itemize}
\item \texttt{0x0005} (\texttt{status\_request})
\item \texttt{0x0010} (\texttt{supported\_groups})
\item \texttt{0x3374} (\texttt{next\_protocol\_negotiation})
\item \texttt{0x0017} (\texttt{extended\_master\_secret})
\end{itemize}

The client's public key length, taken from the \texttt{client key exchange} message, has discriminatory power. As illustrated in Figure \ref{figure:malware_versus_benign_client}, most of the enterprise traffic used a 512-bit (\texttt{ECDHE\_RSA}) public key. In contrast, malware almost exclusively used a 2048-bit (\texttt{DHE\_RSA}) public key.

Finally, we mapped the TLS client parameters to well known client programs that use specific TLS libraries and configurations \cite{ssllabs}. This information could be spoofed, but we feel that this is still a valuable and compact way to represent a client. As shown in Figure \ref{figure:malware_versus_benign_client}, the most popular clients for malware and enterprise TLS connections are quite distinct. In the enterprise setting, we found that the four most common client configurations resembled the most recent releases of the four most popular browsers: \texttt{Firefox 47}, \texttt{Chrome 51}, \texttt{Internet Explorer 11}, and \texttt{Safari 9}. On the other hand, malware most frequently used TLS client parameters that matched those of \texttt{Opera 12}, \texttt{Firefox 46}, and \texttt{Tor 0.2.x}.

\subsubsection{Malware Families}

Table \ref{table:malware_tls_client} gives the most popular TLS client parameters for each of the 18 malware families we had access to. The most popular TLS client was \texttt{Internet Explorer 8}, which was used most frequently by 10 of the 18 families. These families and client values are listed for completeness, but should more accurately be read as utilizing the TLS library provided by the underlying Windows environment.

The Tor client and browser were very popular among the malware families, being the most popular with Deshacop, Dynamer, Razy, Skeeyah, and Toga. Dynamer, Skeeyah, and Symmi all used a 512-bit (\texttt{ECDHE\_RSA}) public key as opposed to the most popular public key: 2048-bit (\texttt{RSA}), which is most likely an artifact of the underlying Windows environment.

Table \ref{table:malware_tls_client} also lists the number of distinct ciphersuite offer vectors observed for each malware family. In this context, a client is taken to be unique if it has a different set of offered ciphersuites and advertised extensions. Some families have very few unique clients, e.g., Bergat. On the other hand, Sality has a large number of distinct ciphersuite offer vectors. And while Sality's most used TLS client offered parameters similar to \texttt{Internet Explorer 8}, it had hundreds of other unique combinations of offered ciphersuites and advertised extensions.

\subsection{TLS Servers}

\subsubsection{Malware versus Enterprise}

Figure \ref{figure:malware_versus_benign_server} illustrates the differences between the servers connected to by the malware and the enterprise TLS clients after filtering clients that used typical Windows XP ciphersuite lists. The filtering was done for the server statistics because those clients have a significant impact on what is sent in the \texttt{server hello} message.

As seen in Figure \ref{figure:malware_versus_benign_server}, the selected ciphersuites of the \texttt{server hello} messages are sharply divided for the majority of enterprise and malicious TLS sessions. The following four ciphersuites were selected by $\sim$90\% of the servers that malware communicated with:
\begin{itemize}
\item \texttt{0x000a} \newline(\texttt{TLS\_RSA\_WITH\_3DES\_EDE\_CBC\_SHA})
\item \texttt{0x0004} (\texttt{TLS\_RSA\_WITH\_RC4\_128\_MD5})
\item \texttt{0x006b} \newline(\texttt{TLS\_DHE\_RSA\_WITH\_AES\_256\_CBC\_SHA256})
\item \texttt{0x0005} (\texttt{TLS\_RSA\_WITH\_RC4\_128\_SHA})
\end{itemize}
These ciphersuites were rarely selected by servers that enterprise hosts communicated with. \texttt{TLS\_RSA\_WITH} \texttt{\_RC4\_128\_MD5} and \texttt{TLS\_RSA\_WITH\_RC4\_128\_SHA} are considered weak.

As one would expect given the lack of advertised TLS extensions by the malware clients, the servers that malware communicated with rarely selected TLS extensions. On the other hand, the servers that the enterprise hosts communicated with had a much greater diversity in the selected TLS extensions with \texttt{0xff01} (\texttt{renegotiation\_info}) and \texttt{0x000b} (\texttt{ec\_point\_formats}) being the most frequent.

We also analyzed information from the servers' certificates. As anticipated, we found that enterprise endpoints most frequently connected to servers with the following certificate subjects:
\begin{itemize}
\item \texttt{*.google.com}
\item \texttt{api.twitter.com}
\item \texttt{*.icloud.com}
\item \texttt{*.g.doubleclick.net}
\item \texttt{*.facebook.com}
\end{itemize}
This distribution of certificate subjects was very long tailed. The certificate subjects of servers that the malware samples communicated with also had a long tail. These certificates were mostly composed of subjects that had characteristics of a domain generation algorithm (DGA) \cite{antonakakis2012throw}, e.g., \texttt{www.33mhwt2j.net}. Although malware mostly communicated with servers that had suspicious certificate subjects, it is also clear that malware communicates with many inherently benign servers, e.g., \texttt{google.com} for connectivity checks or \texttt{twitter.com} for command and control. The following certificate subjects were the most frequent for TLS flows initiated by malware:
\begin{itemize}
\item \texttt{block.io}
\item \texttt{*.wpengine.com}
\item \texttt{*.criteo.com}
\item \texttt{baidu.com}
\item \texttt{*.google.com}
\end{itemize}
Because the DGA-like certificate subjects are counted as unique, they do not show up in this list.

Figure \ref{figure:malware_versus_benign_server} highlights two other interesting features associated with server certificates: the validity of the certificate in days and the number of \texttt{subjectAltName} entries. Interestingly, the high prevalence of connections to \texttt{block.io}, a Bitcoin wallet, heavily skewed the validity (375 days) and number of \texttt{subjectAltName} entries (3) for the certificates of servers that malware connected to.

It is also interesting to note the frequency of TLS servers using certificates that are self-signed. In the enterprise data, 1,352 out of the 1,500,005 TLS sessions, or $\sim$0.09\%, used a self-signed certificate. In the malware data, 947 out of the 133,744 TLS sessions, or $\sim$0.7\%, used a self-signed certificate, which is roughly an order of magnitude more frequent than the enterprise case.

\begin{table*}[t!]\small
\center
  \begin{tabular}{l|r|r|r|r}
  	\hline
  	& \multicolumn{2}{|c|}{All Data} & \multicolumn{2}{|c}{No \texttt{SChannel}} \\
    Dataset & Total Accuracy & 0.01\% FDR & Total Accuracy & 0.01\% FDR \\
    \hline
    \hline
    Meta+SPLT+BD+TLS+SS & 99.6\% & 92.6\% & 99.6\% & 87.4\% \\
    \hline
    Meta+SPLT+BD+TLS & 99.6\% & 92.8\% & 99.6\% & 87.2\% \\
    \hline
    TLS & 98.2\% & 63.8\% & 96.7\% & 59.1\% \\
    \hline
    Meta+SPLT+BD & 98.9\% & 1.3\% & 98.5\% & 0.9\% \\
    \hline
  \end{tabular}
  \vspace{2mm}
  \caption{Classifier accuracy for different combinations of data features, showing the overall accuracy and the accuracy at a 0.01\% FDR.}
  \label{table:classification_results}
\end{table*}

\begin{table*}[t!]\small%\normalsize
\center
  \begin{tabular}{l|r|r|r|r}
  	\hline
%    Malware Family & Meta+SPLT+BD & TLS Only & Meta+SPLT+BD+TLS & All+SS\\
    \hspace{1.5mm}Malware Family\hspace{1.5mm} & \hspace{1.5mm}Meta+SPLT\hspace{1.5mm} & \hspace{1.5mm}TLS Only\hspace{1.5mm} & \hspace{1.5mm}Meta+SPLT\hspace{1.5mm} & \hspace{1.5mm}All+SS\hspace{1.5mm}\\
     & +BD\hspace{7mm} &  & +BD+TLS\hspace{3mm} & \\
    \hline
    \hline
    Bergat* & 100.0\% & 100.0\% & 100.0\% & 100.0\% \\
    \hline
    Kazy* & 98.5\% & 99.5\% & 99.8\% & 100.0\% \\
    \hline
    Parite* & 99.3\% & 97.8\% & 99.6\% & 99.6\% \\
    \hline
    Sality* & 95.0\% & 94.1\% & 97.7\% & 98.0\% \\
    \hline
    Tescrypt* & 89.8\% & 95.6\% & 97.6\% & 97.6\% \\
    \hline
    Upatre* & 99.9\% & 98.7\% & 100.0\% & 100.0\% \\
    \hline
    Virtob* & 99.2\% & 98.8\% & 99.4\% & 99.4\% \\
    \hline
    Yakes* & 88.7\% & 98.5\% & 99.7\% & 99.7\% \\
    \hline
    Zbot* & 98.9\% & 99.6\% & 99.7\% & 100.0\% \\
    \hline
    Zusy* & 98.6\% & 88.7\% & 99.9\% & 99.9\% \\
    \hline
    \hline
    Deshacop & 93.0\% & 63.6\% & 96.1\% & 96.1\% \\
    \hline
    Dridex & 16.5\% & 68.7\% & 78.5\% & 97.9\% \\
    \hline
    Dynamer & 95.4\% & 78.8\% & 95.7\% & 96.5\% \\
    \hline
    Razy & 91.5\% & 77.1\% & 95.9\% & 96.8\% \\
    \hline
    Skeeyah & 95.9\% & 82.1\% & 98.6\% & 98.6\% \\
    \hline
    Symmi & 99.1\% & 92.4\% & 99.8\% & 99.8\% \\
    \hline
    Toga & 100.0\% & 100.0\% & 100.0\% & 100.0\% \\
    \hline
    Virlock & 100.0\% & 100.0\% & 100.0\% & 100.0\% \\
    \hline   
  \end{tabular}
  \vspace{2mm}
  \caption{Classifier accuracy when separated by family. Families with an (*) offered an ordered ciphersuite list that matched a list found in the default Windows XP \texttt{SChannel} implementation. Malware data from August and September 2015, and enterprise data from May and June 2016 were used for training; these malware samples were collected from October 2015 to May 2016. Results using unencrypted TLS handshake messages are given in addition to results based on only standard side-channel features. The two baselines are the first two data columns: side-channel-only and TLS-only.}
  \label{table:malware_classification}
\end{table*}

\subsubsection{Malware Families}

Table \ref{table:malware_tls_server} lists several interesting statistics about the servers that malware most often connects to. Some of the malicious families connect to a large number of unique IP addresses, e.g., Symmi and Dynamer. The family with the most flows, Virlock, only connects to 1 unique IP address owned by \texttt{block.io}.

We observed 10 families that made use of self-signed certificates. ZBot was the most frequent offender, with the subject of these certificates being \texttt{tridayacipta.com}, a domain name that has many detections on VirusTotal \cite{virustotal}.

Common certificate subjects also allow one to make inferences about the tools that the malware families use and the functionality that the families support. For instance, Deshacop and Tescrypt have \texttt{*.onion.to} as the most common certificate subject, and, as anticipated, both have many samples that have TLS client configurations that indicate that they use the \texttt{Tor Browser}. The \texttt{Tor Browser} is the most prevalent client for Deshacop, and the second most prevalent client for Tescrypt. Symmi's most common certificate subject is \texttt{*.criteo.com}, an ad service. This could indicate Symmi's intent to perform click-fraud.

\section{Classifying Encrypted Traffic}
\label{sec:malware_classification}

We used a logistic regression classifier with an $l1$ penalty \cite{koh2007interior} for all classification results. For the initial binary-class classification results, we trained four machine learning classifiers using different subsets of data features we collected. The first classifier used the flow metadata (Meta), the sequence of packet lengths and inter-arrival times (SPLT), and the distribution of bytes (BD). The second classifier only used the TLS information (TLS). The third classifier was trained using the same features as the first, with the addition of the TLS client information, specifically, the offered ciphersuites, advertised extensions, and the client's public key length. The fourth classifier was trained with all data, and an additional, custom feature: whether the server certificate was self-signed (SS).

\subsection{Malware versus Enterprise}

To demonstrate the value of the additional TLS features in a classification setting, we use 10-fold cross-validation and all of the malicious TLS flows collected from August 2015 until May 2016, and a random subset of the May and June 2016 enterprise network's TLS flows. In total, there were 225,740 malicious and 225,000 enterprise flows for this experiment. To account for the bias that the Windows XP-based sandbox could introduce, we also present results on a dataset composed of only flows that did not offer an ordered ciphersuite list that matched a list found in the default Windows XP \texttt{SChannel} implementation \cite{ssllabs}: 133,744 malicious and 135,000 enterprise TLS flows.

The 10-fold cross-validation results for the above problem is shown in Table \ref{table:classification_results}. We see that using all available data views significantly improves the results. A 1-in-10,000 false discovery rate (FDR) is defined as the accuracy on the positive class given that only 1 false positive is allowed for every 10,000 true positives. As these results show, not using TLS header information leads to significantly worse performance, especially in the important case of a fixed, 1-in-10,000 FDR. The removal of the Windows XP \texttt{SChannel} TLS flows had no effect on the total accuracy of the classifiers based on all data views, but does reduce the performance at a 1-in-10,000 FDR by $\sim$5\%.

%\begin{enumerate}
%\item Meta+SPLT+BD: 97.24\% total accuracy and 0.10\% 1-in-10,000 FDR
%\item TLS: 98.52\% total accuracy and 61.97\% 1-in-10,000 FDR
%\item Meta+SPLT+BD+TLS: 99.48\% total accuracy and 90.49\% 1-in-10,000 FDR
%\item ALL+SS: 99.49\% total accuracy and 90.78\% 1-in-10,000 FDR
%\end{enumerate}

\begin{figure*}[t!]
	\centering
   \includegraphics[scale=0.74]{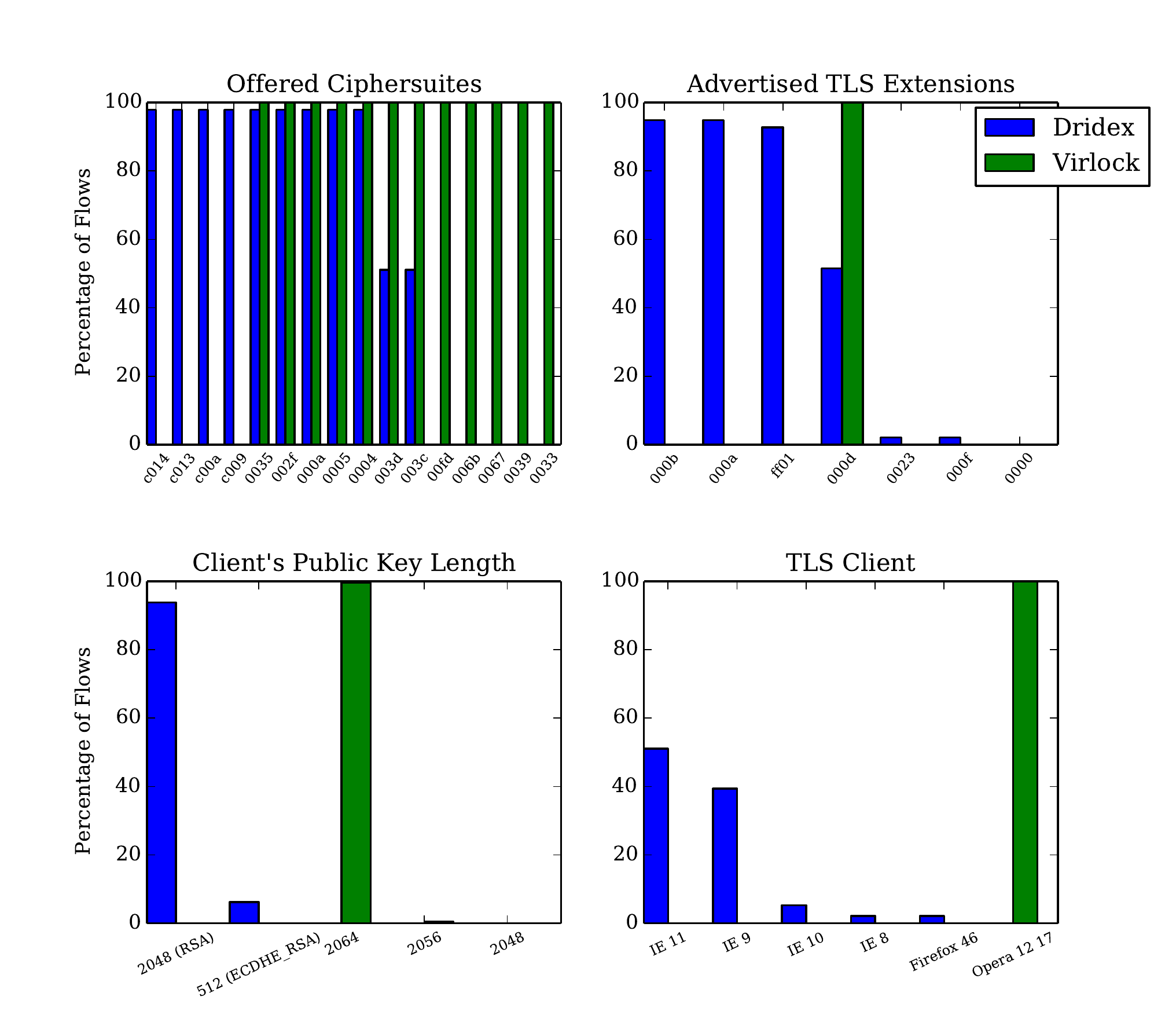}
   \caption{Dridex's use of TLS versus that of Virlock's. Some values and the full ciphersuite names were omitted for clarity of presentation. Ciphersuites and extensions are represented as hex codes, which are given in full in Appendix \ref{app:hex_codes}.}
   \label{figure:dridex_vs_virlock}
\end{figure*}

\subsection{Malware Families}

To determine how well a trained classifier is able to detect the TLS flows generated by the different malware families, we first trained the four classifiers from Table \ref{table:classification_results} on the same 225,000 enterprise flows as above for the negative class, and 76,760 malicious TLS flows collected during August and September 2015 for the positive class. These binary classifiers were applied to the testing data consisting of the TLS flows from October 2015 to May 2016, summarized in Table \ref{table:malware_summary}. While we do not remove flows that offered an ordered ciphersuite list that matched a list found in the default Windows XP \texttt{SChannel} implementation in this experiment, we do make explicit the families that have this bias in the majority of their flows.

Table \ref{table:malware_classification} lists the classification accuracy of the four classifiers for each family. Because only malware data was used to test the trained classifiers, false positives for this experiment are ill-defined and are therefore not reported. In the August and September 2015 malware training data, there was strong representation of the malicious families presented in this paper, but there were not any exact SHA1 matches. There were four families that had no representation in August or September: Bergat, Yakes, Razy, and Dridex.

For the most part, combining traditional flow metadata, typical side-channel information, and the TLS specific features led to the best performing machine learning models. Out of all families, our classifiers with all data views performed the worst on Deshacop with a 96.1\% true positive rate. With respect to only the malware families that primarily used ciphersuites similar to those used by Windows XP \texttt{SChannel}-based clients, our classifiers with all data views performed the worst on Tescrypt with a 97.6\% true positive rate. Both of these families most often visited servers with a server certificate subject of \texttt{*.onion.to}, and use TLS client configurations that indicate the \texttt{Tor Browser} for some of their TLS connections. This is particularly interesting because a major goal of the \texttt{Tor Browser} is to maintain the privacy of its users, which in this case are the malware families.

The classifier based only on the TLS data was able to perform quite well on the malware families that used TLS client configurations that matched those of Windows XP \texttt{SChannel}-based clients, but this result is not guaranteed to hold if the malware runs on another operating system. The TLS-only classifier performed the worst on most of the families that used TLS client configurations that did not match those of Windows XP \texttt{SChannel}-based clients, with the exception of Toga and Virlock. Both of these families did a poor job at varying the TLS client parameters in our dataset, and they both used TLS client parameters that indicated older versions of clients: Toga $\rightarrow$ \texttt{Tor 0.2.2} and Virlock $\rightarrow$ \texttt{Opera 12}.

\begin{figure*}[t!]
	\centering
   \includegraphics[scale=0.59]{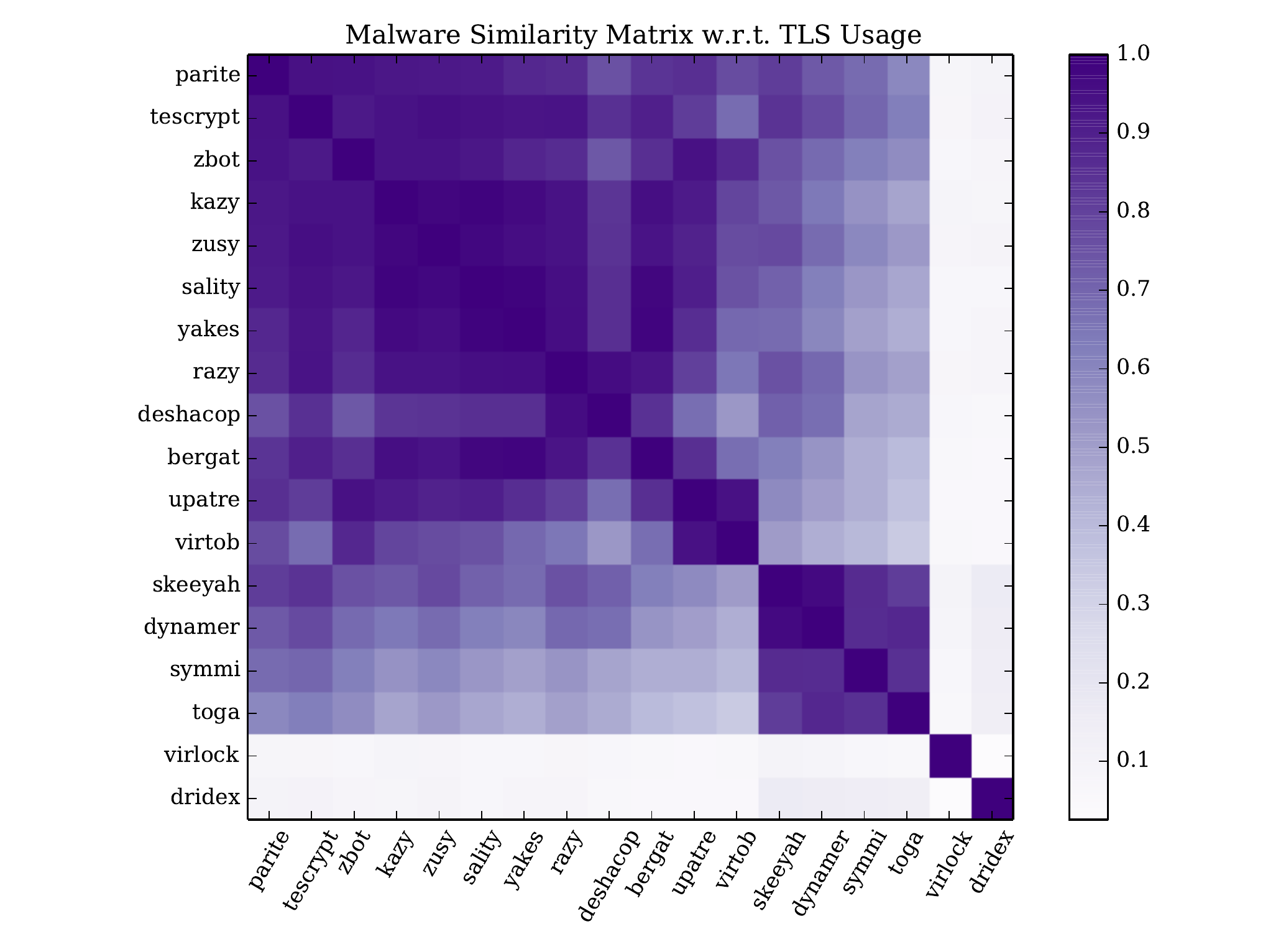}
   \caption{Similarity Matrix for the different malware families with respect to the observed TLS client's parameters.}
   \label{figure:malware_similarity_matrix}
\end{figure*}

The machine learning classifiers were able to perform reasonably on most malware families, with the exception of Dridex. Dridex was one of four families that did not have any representation in the training data. The classifier on the other three families, Bergat, Yakes, and Razy, had $\sim$96-100.0\% total accuracy. In the case of Bergat and Yakes, this good performance is expected because these families offered an ordered ciphersuite list that matched a list found in the default Windows XP \texttt{SChannel} implementation.

Figure \ref{figure:dridex_vs_virlock} shows Dridex's use of TLS from a client point-of-view. Unlike most of the other families, Dridex most often selects:
\begin{itemize}
\item \texttt{0x002f} (\texttt{TLS\_RSA\_WITH\_AES\_128\_CBC\_SHA})
\end{itemize}
Figure \ref{figure:malware_versus_benign_server} shows that this ciphersuite is not uncommon for enterprise TLS sessions. Dridex also advertises several TLS extensions and offers many current ciphersuites in the \texttt{client hello} message.

Figure \ref{figure:dridex_vs_virlock} also compares Dridex's TLS usage with that of Virlock's. Virlock is an example of a malicious family that used the same TLS client for every sample that we observed, and was able to be easily classified, i.e., all four classifiers achieved 100\% accuracy. While Dridex offers a variety of strong ciphersuites, Virlock offers a smaller set of outdated ciphersuites. Virlock also only  advertises the \texttt{signature\_algorithms} TLS extensions. Another significant difference between these two families is that Virlock did not alter its TLS client's behavior once in our entire dataset. Virlock always used the same client parameters that are similar to those of Opera 12. Virlock's lack of adaptation makes it trivial for a machine learning, or a rule-based, system to classify. Dridex's use of multiple TLS clients made a significant difference in terms of detection efficacy.

\begin{figure*}[t!]
	\centering
   \includegraphics[scale=0.54]{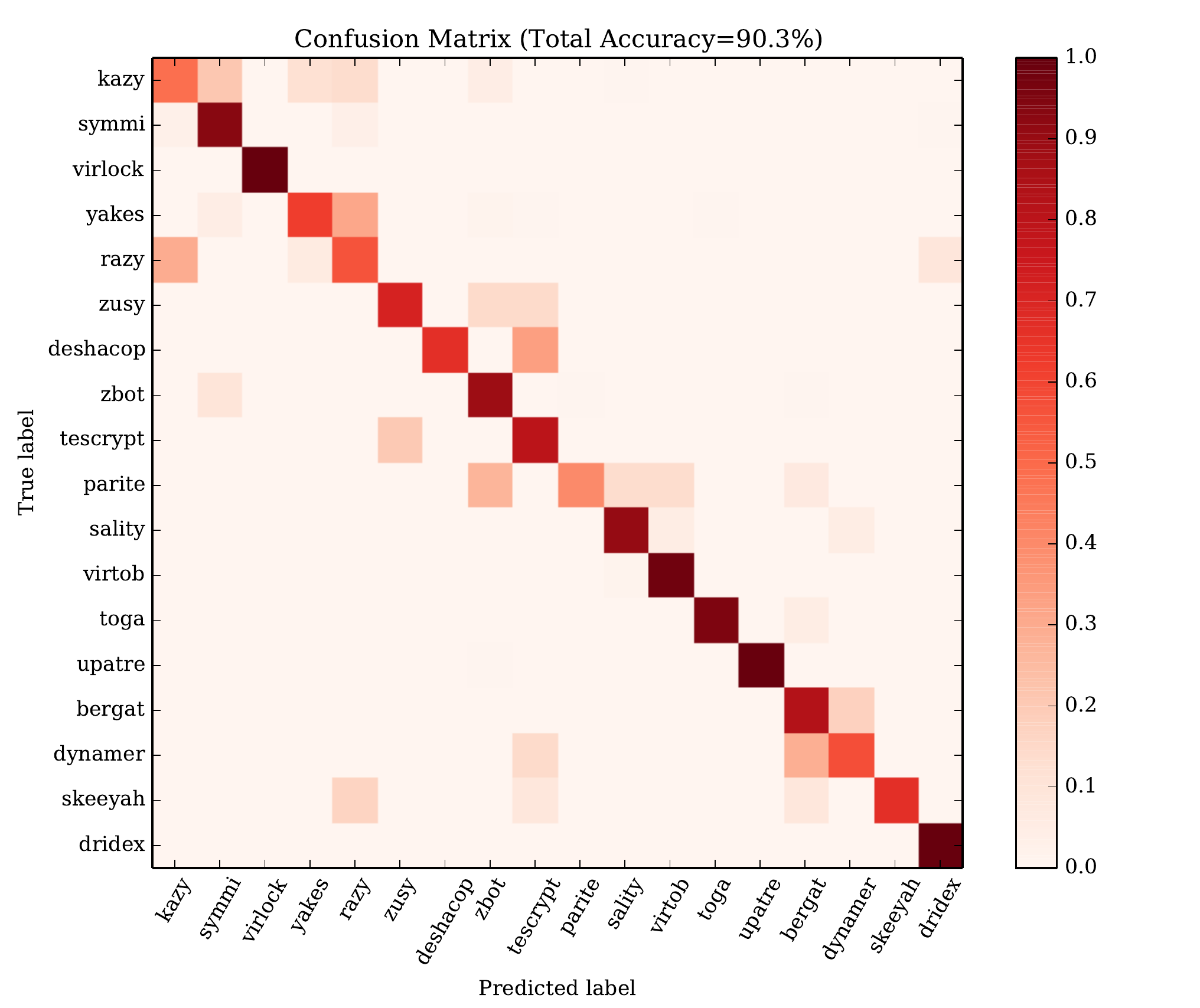}
   \caption{Confusion matrix for the 18-class malware family classifier. The total 10-fold accuracy of the machine learning model was 90.3\%.}
   \label{figure:confusion_matrix}
\end{figure*}

As we now show, awareness of self-signed certificates proved to be crucial. The classification of Dridex using Meta+SPLT+BD+TLS, 78.5\%, does not inspire confidence in a system designed to detect malicious, encrypted traffic. Our hypothesis was that, although Dridex varies the behavior of its TLS clients, there might be an invariant with the servers that Dridex communicates with that would allow us to more easily classify these encrypted flows. Upon manual inspection, this hypothesis was confirmed. We included a binary feature indicating whether the server certificate was self-signed (denoted as SS), and retrained our machine learning classifier with this new feature. The 10-fold cross-validation results on the training data were nearly identical. With the self-signed feature, the new classifier with all data sources achieved an accuracy of 97.9\% on Dridex, a significant improvement.

\section{Family Attribution}
\label{sec:family_attribution}

Being able to accurately attribute malware samples to a known family is highly valuable. Attribution provides incident responders with actionable prior information before they begin to reverse engineer malware samples. From a network point-of-view, this attribution can help to prioritize the incident responders time, i.e., available resources should be assigned to investigate more serious infections. In these results, there are no enterprise samples; we only consider malicious samples and their associated families.

To analyze the differences between the TLS parameters used by different malware families, we used the malware samples from October 2015 to May 2016 that had an identifiable family name as described in Section \ref{sec:data}. This process pruned our original set of 20,548 samples to 5,623 unique samples across 18 families. These samples generated 25,793 TLS encrypted flows that successfully negotiated the full TLS handshake and sent application data.

\subsection{Similar TLS Usage}

Figure \ref{figure:malware_similarity_matrix} shows a similarity matrix for the 18 malware families with respect to their TLS clients. The offered ciphersuites, advertised extensions, and the client's public key length were used as features, and a standard squared exponential similarity function was used to compute the similarity values:
\begin{equation}
exp\left(-\lambda \sum_{i,j} (x_i - x_j)^2\right)
\end{equation}
with $\lambda = 1$, and $x_i$ being the mean of the feature vectors for the $i$'th family. The diagonal of this matrix will be 1.0 because each family will be perfectly self-similar.

There is a lot of structure in Figure \ref{figure:malware_similarity_matrix}. The upper left block consists of families that have some number of flows that use the default Windows XP TLS library. The group of Skeeyah, Dynamer, Symmi, and Toga all heavily make use of offered ciphersuite lists and advertised extensions that are indicative of \texttt{Tor 0.2.x}. Dridex and Virlock were the two most dissimilar malware families. And while Dridex was difficult to accurately classify, Virlock was trivial. Uniqueness is not always a desirable quality.

\subsection{Multi-Class Classification}

Finally, to assess the malware family attribution potential of TLS flows, we used the data listed in Table \ref{table:malware_summary}, and did not remove samples that offered an ordered ciphersuite list that matched a list found in the default Windows XP \texttt{SChannel} implementation in this experiment because all of the samples would have the same bias. We position the problem of attributing a malicious TLS flow to a known malware family as a multi-class classification problem. For this analysis, we use all of the malware families and data features described in Section \ref{sec:data}. Similar to the enterprise versus malware results in Section \ref{sec:malware_classification}, we used 10-fold cross validation and $l1-$ multinomial logistic regression \cite{krishnapuram2005sparse}. We not only present our results in terms of overall classification accuracy, but also as a confusion matrix showing the true positives and false positives broken down per-family. This was done to illustrate that we were not simply using a na\"ive majority-class classifier, but were in fact making useful inferences.

Using all available data features led to the best cross-validated performance, with a total accuracy of 90.3\% for the 18-class classification problem using a single, encrypted flow. The confusion matrix for this problem is shown in Figure \ref{figure:confusion_matrix}. For a given row (family) in the confusion matrix, the column entries represent the percentage of samples identified as that specific family. A perfect confusion matrix would have all of its weight focused on the diagonal. As an example, most of Kazy's TLS flows, the first row, were identified as Kazy, the first column. Some of Kazy's TLS flows were also identified as Symmi (column: 2), Yakes (column: 4), Razy (column: 5), and Zbot (column: 8).

%\begin{figure*}[t!]
%	\centering
%   \includegraphics[scale=0.55]{figures/yakes_vs_razy.pdf}
%   \caption{Yakes's use of TLS versus that of Razy's. Some specific ciphersuite and extension values were omitted for clarity of presentation. Ciphersuites and extensions are represented as hex codes and the client's public key length is given in bits.}
%   \label{figure:yakes_versus_razy}
%\end{figure*}

The majority of the TLS flows were attributed to the appropriate family with $\sim$80-90\% accuracy. Again, the two exceptions are Dridex and Virlock. Attribution for these two families are trivial, in large part because of their distinctive use of TLS compared to other malicious families.

There were two sets of two families that the multiclass classification algorithm had problems differentiating. The first of these was Bergat and Dynamer. Interestingly, Bergat used a Windows XP \texttt{SChannel}-like TLS client, but Dynamer used a \texttt{tor 0.2.2}-like TLS client. The confusion came from the other data views, specifically the sequence of packet lengths. Both of these families most often connected to servers at \texttt{www.dropbox.com}, and had similar communication patterns.

Finally, Yakes and Razy were another two malicious families that the multi-class classifier could not differentiate. Like Bergat and Dynamer, Yakes and Razy most often connected to servers at \texttt{baidu.com}. In fact, these two families are subfamilies of the Ramnit family. Upon manual inspection, the network behavior of Yakes and Razy looked mostly identical.

%\begin{figure*}[t!]
%	\centering
%   \includegraphics[scale=0.65]{figures/confusion_matrix_multi.pdf}
%   \caption{Confusion matrix for the 18-class classification problem using a multiple flow classification model. The total 10-fold accuracy of the machine learning model was 91.3\%.}
%   \label{figure:confusion_matrix_multi}
%\end{figure*}

%\subsubsection{Incorporating Multiple Flows}

%Determining the malware family based on a single, encrypted flow is an unnecessarily difficult problem. In our dataset, the malware samples often created many encrypted flows that can be used for attribution. In this framework, one could initially classify all of the flows in a 5 minute sliding window for a given host, and use the suspicious flows to perform family attribution. We first trained an independent flow, multi-class classifier. Then, for each window in the testing set, each flow was classified, and a majority vote was used to classify all flows within the window. This is similar to ensemble methods in machine learning \cite{dietterich2000ensemble}. Figure \ref{figure:confusion_matrix_multi} shows the confusion matrix resulting from 10-fold cross validation on this problem. The accuracy of the multi-class problem increased from 90.3\% using single, encrypted flows to 93.2\% using a simplistic multiple flow algorithm. Comparing to Figure \ref{figure:confusion_matrix}, it is easy to see that this scheme increased the accuracy of most families, notably Yakes and Razy.

Determining the malware family based on a single, encrypted flow is an unnecessarily difficult problem. In our dataset, the malware samples often created many encrypted flows that can be used for attribution. In this framework, one could initially classify all of the flows in a 5 minute sliding window for a given host, and use the suspicious flows to perform family attribution. We first trained an independent flow, multi-class classifier. Then, for each window in the testing set, each flow was classified, and a majority vote was used to classify all flows within the window. This is similar to ensemble methods in machine learning \cite{dietterich2000ensemble}. The confusion matrix resulting from 10-fold cross validation on this problem looked very similar to that shown in Figure \ref{figure:confusion_matrix}. The accuracy of the multi-class problem increased from 90.3\% using single, encrypted flows to 93.2\% using a simplistic multiple flow algorithm. While there were several families that had improved performance, this simple, multi-flow scheme increased the accuracy of Yakes and Razy most notably. This was most likely because Razy was more promiscuous.

\section{Related Work}
\label{sec:related_work}

Identifying threats in encryption poses significant challenges. Nevertheless, the security community has put forth two solutions to solve this problem. The first involves decrypting all traffic that flows through a security appliance: Man-in-the-Middle (MITM) \cite{CallegatiMITM09}. Once the traffic has been decrypted, traditional signature-based methods, such as Snort \cite{snort}, can be applied. While this approach can be successful at finding threats, there are several important shortcomings. First, this method does not respect the privacy of the users on the network. Second, this method is computationally expensive and difficult to deploy and maintain. Third, this method relies on malware clients and servers to not change their behavior when a MITM interposes itself.

The second method of identifying threats in encrypted network traffic leverages flow-based metadata. These methods examine high-level features of a network flow, such as the number of packets and bytes within a flow. This data is typically exported and stored as IPFIX \cite{ipfix} or NetFlow \cite{netflow}. There have been several papers that push the limits of traditional flow monitoring systems. For instance, \cite{bilge2012disclosure} uses NetFlow and external reputation scores to classify botnet traffic. This work can also be applied to encrypted network traffic, but does not take advantage of the TLS-specific data features.

In addition to pure flow-based features to detect malware's network traffic, there has been many papers that augment this data with more detailed features about a flow \cite{dietrich2013cocospot,gu2008botminer,moore2005internet,wang2006anomalous,wang2015seeing,Williams06apreliminary,wurzinger2009automatically,zander2005automated}.  This work can been seen as utilizing side-channel attacks, such as analyzing the sizes and inter-arrival times of packets, to learn more information about a flow. In \cite{panchenko2016website}, the authors derive features based on the packet sizes to perform website fingerprinting attacks against encrypted traffic. In our work, we are only concerned with identifying malware communication and we use information specific to the TLS protocol.

There has been previous work that uses active probing \cite{durumeric2013zmap} and passive monitoring to gain visibility into how TLS is used in the wild \cite{holz2015tls}. Unlike \cite{holz2015tls}, our results specifically highlight malware's use of the TLS protocol, and show how data features from TLS can be used in rules and classifiers.

Malware clustering and family attribution has had a lot of exposure in the academic literature \cite{anderson2012multiple,bayer2009scalable,perdisci2010behavioral,rieck2008learning}.  This work has taken a variety of data source, e.g, \texttt{HTTP} or dynamic system call traces, and clustered the samples to attribute a sample to a malicious family. In contrast, our work gives an in-depth analysis of how malware uses TLS, and shows how data features from passive monitoring of TLS can be used for accurate malware identification and family attribution.

\section{Limitations and Future Work}
\label{sec:limitations}

Our method for collecting malware data was straightforward and allowed us to quickly generate a large volume of network data, but the dependence on Windows XP and 5 minute runs introduced some biases in our presented results. We accounted for these biases by specifically considering the cases in which the TLS features reflected the operating system and not the malware, and either analyzing the data with those cases removed, or clearly labeling and analyzing those cases otherwise. Accounting for the bias caused by the sandbox was essential to understanding the actual malware use of TLS. From a practitioner’s point of view, however, it is sometimes worthwhile to consider the raw, biased data. Malware often targets obsolete and unpatched software because it is vulnerable, and thus it is biased in the same direction as the sandbox. We leave running these samples under multiple environments and collecting the additional results for future work.

%Server certificate can be obfuscated, could do active probing

After family names were associated with our malware samples, the original set of 20,548 samples that used TLS was reduced to a set of 5,623 unique samples across 18 families. It is difficult to reliably determine the family, if any, associated with a malware sample, even in a structured sandbox setting. While our multi-class, malware family classifier can reasonably be criticized for failing to provide attribution for $\sim$3/4 of the malware samples, this fact reflects the difficulty of family attribution in a dynamic analysis environment, and not a limitation of the underlying approach. In future work, the malware families for the training data can be determined by a robust clustering algorithm \cite{anderson2012multiple} instead of relying on a consensus vote from VirusTotal \cite{virustotal}.

Like nearly all other methods of threat detection, a motivated threat actor could attempt to evade detection by mimicking the features of enterprise traffic. For instance, in our case, this could take the form of attempting to offer the same TLS parameters as a popular \texttt{Firefox} browser and using a certificate issued by a reputable certificate authority. But, while evasion is always possible in principle, in practice it poses challenges for the malware operator. Mimicking a popular HTTPS client implementation requires an ongoing and non-trivial software engineering effort; if a client offers a TLS ciphersuite or extension that it cannot actually support, the session is unlikely to complete. On the server side, the certificate must mimic the issuer, subjectAltName, time of issuance, and validity period of the benign server. In either case, the detection methods outlined in this paper are not meant to be exhaustive, and in a robust system, these methods would only be one facet of the final solution. An example of extending this methodology for robustness would be to build a profile for an endpoint based on the \texttt{user-agent} string advertised in the unencrypted HTTP flows. If the TLS parameters indicate a user agent that has not been observed on an endpoint, this could be an interesting indicator of compromise.

All of the classification results presented in this paper used 10-fold cross-validation and $l1$-logistic regression. We have found this classifier to be very efficient and to perform extremely well for network data feature classification. This model reports a probabilistic output, allowing one to easily change the threshold of the classifier. We did compare $l1$-logistic regression with a support vector machine (Gaussian kernel, width adjusted through CV), and found \textbf{no} statistically-significant improvement using a 10-fold paired $t$-test at a 5\% significance level \cite{dietterich1998approximate}. Because of the added computational resources needed to train the SVM and the chosen model's robustness against overfitting \cite{yuan2012improved}, we only reported the $l1$-logistic regression results. We leave examining alternative models and quantifying their advantages for future work.

\section{Conclusions}
\label{sec:conclusions}

Understanding malware's use of TLS is imperative for developing appropriate techniques to identify threats and respond to those threats accordingly. In this paper, we reviewed what TLS parameters malware typically uses from both the perspective of the TLS client and the TLS servers that the samples communicated with. Even when we accounted for the bias caused by the underlying sandbox's operating system, we found that malware generally offers and selects weak ciphersuites and does not offer the variety of extensions that we see in enterprise clients.

We also analyzed the TLS usage of malware on a per family basis. We identified malware families that are most likely to use TLS client parameters that matched the TLS library provided by Windows XP, the underlying operating system of the sandbox, e.g., Bergat and Yakes; malware families that use TLS client parameters that matched the TLS library provided by the underlying operating system in addition to hundreds of other TLS client configurations, e.g., Sality; and families that exclusively used TLS client configurations that do not match the TLS libraries supplied by the underlying operating system, e.g., Virlock. As anticipated, we found that families who actively evolve their usage of TLS are more difficult to classify. We also found a malware family that used TLS parameters that are similar to those found on an enterprise network, and was difficult to classify: Dridex. But, if we leverage additional, domain-specific knowledge such as whether the TLS certificate was self-signed, we can significantly increase the performance of our classifiers.

We showed that the differences in how malware families use TLS can be used to attribute malicious, encrypted network flows to a specific malware family. We also observed some malware families using TLS in exactly the same way, e.g., Yakes and Kazy, which most often offered an ordered ciphersuite list that matched a list found in the default Windows XP \texttt{SChannel} implementation. We demonstrated an accuracy of 90.3\% for the family attribution problem when restricted to a single, encrypted flow, and an accuracy of 93.2\% when we made use of all encrypted flows within a 5-minute window.

We conclude that data features that are passively observed in TLS provide information about both the client and server software and its configuration. This data can be used to detect malware and perform family attribution, either through rules or classifiers. Malware's TLS data features obtained from sandboxes are biased, and it is essential to understand and account for this bias when using these features

%\romanappendices

\bibliographystyle{splncs03}
\bibliography{raid.bib}

\appendices

\section{Ciphersuite and Extension Hex Codes}
\label{app:hex_codes}

\begin{table}[b!]\small%\normalsize
\center
  \begin{tabular}{l|r}
  	\hline
    \hspace{.5mm}Hex Code\hspace{.5mm} & \hspace{.5mm}Ciphersuite\hspace{.5mm} \\
    \hline
    \hline
    \texttt{0x0004} & \texttt{TLS\_RSA\_WITH\_RC4\_128\_MD5} \\
    \hline
    \texttt{0x0005} & \texttt{TLS\_RSA\_WITH\_RC4\_128\_SHA} \\
    \hline
    \texttt{0x000a} & \texttt{TLS\_RSA\_WITH\_3DES\_EDE\_CBC\_SHA} \\
    \hline
    \texttt{0x002f} & \texttt{TLS\_RSA\_WITH\_AES\_128\_CBC\_SHA} \\
    \hline
    \texttt{0x0033} & \texttt{TLS\_DHE\_RSA\_WITH\_AES\_128\_CBC\_SHA} \\
    \hline
    \texttt{0x0035} & \texttt{TLS\_RSA\_WITH\_AES\_256\_CBC\_SHA} \\
    \hline
    \texttt{0x0039} & \texttt{TLS\_DHE\_RSA\_WITH\_AES\_256\_CBC\_SHA} \\
    \hline
    \texttt{0x003c} & \texttt{TLS\_RSA\_WITH\_AES\_128\_CBC\_SHA256} \\
    \hline
    \texttt{0x003d} & \texttt{TLS\_RSA\_WITH\_AES\_256\_CBC\_SHA256} \\
    \hline
    \texttt{0x0067} & \texttt{TLS\_DHE\_RSA\_WITH\_AES\_128\_CBC\_} \\
     & \texttt{SHA256} \\
    \hline
    \texttt{0x006b} & \texttt{TLS\_DHE\_RSA\_WITH\_AES\_256\_CBC\_} \\
     & \texttt{SHA256} \\
    \hline
    \texttt{0x00fd} & \texttt{unassigned} \\
    \hline
    \texttt{0xc009} & \texttt{TLS\_ECDHE\_ECDSA\_WITH\_AES\_128\_CBC\_} \\
     & \texttt{SHA} \\
    \hline
    \texttt{0xc00a} & \texttt{TLS\_ECDHE\_ECDSA\_WITH\_AES\_256\_CBC\_} \\
     & \texttt{SHA} \\
    \hline
    \texttt{0xc013} & \texttt{TLS\_ECDHE\_RSA\_WITH\_AES\_128\_CBC\_SHA} \\
    \hline
    \texttt{0xc014} & \texttt{TLS\_ECDHE\_RSA\_WITH\_AES\_256\_CBC\_SHA} \\
    \hline
    \texttt{0xc02b} & \texttt{TLS\_ECDHE\_ECDSA\_WITH\_AES\_128\_GCM\_} \\
     & \texttt{SHA256} \\
    \hline
    \texttt{0xc02f} & \texttt{TLS\_ECDHE\_RSA\_WITH\_AES\_128\_GCM\_} \\
     & \texttt{SHA256} \\
    \hline
    \texttt{0xc030} & \texttt{TLS\_ECDHE\_RSA\_WITH\_AES\_256\_GCM\_} \\
     & \texttt{SHA384} \\
    \hline
  \end{tabular}
  \vspace{2mm}
  \caption{Hex code to ciphersuite mapping for ciphersuites used in figures.}
  \label{table:ciphersuites}
\end{table}

\begin{table}[b!]\small%\normalsize
\center
  \begin{tabular}{l|r}
  	\hline
    \hspace{.5mm}Hex Code\hspace{.5mm} & \hspace{.5mm}Ciphersuite\hspace{.5mm} \\
    \hline
    \hline
    \texttt{0x0000} & \texttt{server\_name} \\
    \hline
    \texttt{0x0005} & \texttt{status\_request} \\
    \hline
    \texttt{0x000a} & \texttt{supported\_groups} \\
    \hline
    \texttt{0x000b} & \texttt{ec\_point\_formats} \\
    \hline
    \texttt{0x000d} & \texttt{signature\_algorithms} \\
    \hline
    \texttt{0x000f} & \texttt{heartbeat} \\
    \hline
    \texttt{0x0010} & \texttt{application\_layer\_protocol\_} \\
     & \texttt{negotiation} \\
    \hline
    \texttt{0x0012} & \texttt{signed\_certificate\_timestamp} \\
    \hline
    \texttt{0x0015} & \texttt{padding} \\
    \hline
    \texttt{0x0017} & \texttt{extended\_master\_secret} \\
    \hline
    \texttt{0x0023} & \texttt{SessionTicket TLS} \\
    \hline
    \texttt{0x3374} & \texttt{next\_protocol\_negotiation} \\
    \hline
    \texttt{0x7550} & \texttt{channel\_id} \\
    \hline
    \texttt{0xff01} & \texttt{renegotiation\_info} \\
    \hline
  \end{tabular}
  \vspace{2mm}
  \caption{Hex code to extension mapping for extensions used in figures.}
  \label{table:extensions}
\end{table}

\end{document}